\documentclass[rsi,amsmath,amssymb,preprint]{revtex4-1}
\usepackage{graphicx}		
\usepackage{dcolumn}		
\usepackage{bm}			

\def\wisk#1{\ifmmode{#1}\else{$#1$}\fi}

\def\apj {The Astrophysical Journal}

\begin{document}

\title{The Balloon-Borne Cryogenic Telescope Testbed Mission: \\
Bulk Cryogen Transfer at 40 km Altitude}

\author{A. Kogut}
  \affiliation{Code 665, Goddard Space Flight Center, Greenbelt MD 20771 USA}
  \email{alan.j.kogut@nasa.gov}
\author{S. Denker}
  \affiliation{Department of Mechanical Engineering, University of Maryland, College Park, MD 20742 USA}
\author{N. Bellis}
  \affiliation{Sigma Space Corp, Lanham MD 20706 USA}
  \altaffiliation{Code 665, Goddard Space Flight Center, Greenbelt MD 20771 USA}
\author{T. Essinger-Hileman}
  \affiliation{Code 665, Goddard Space Flight Center, Greenbelt MD 20771 USA}
\author{L. Lowe}
  \affiliation{Sigma Space Corp, Lanham MD 20706 USA}
  \altaffiliation{Code 665, Goddard Space Flight Center, Greenbelt MD 20771 USA}
\author{P.Mirel}
  \affiliation{Sigma Space Corp, Lanham MD 20706 USA}
  \altaffiliation{Code 665, Goddard Space Flight Center, Greenbelt MD 20771 USA}


\begin{abstract}
The  Balloon-Borne Cryogenic Telescope Testbed (BOBCAT) 
is a stratospheric balloon payload
to develop technology for a future cryogenic suborbital observatory.
A series of flights are intended 
to establish ultra-light dewar performance
and open-aperture observing techniques
for large (3--5 meter diameter) cryogenic telescopes
at infrared wavelengths.
An initial flight in 2019 demonstrated bulk transfer
of liquid nitrogen and liquid helium
at stratospheric altitudes.
An 827 kg payload carried
14 liters of liquid nitrogen (LN2)
and 268 liters of liquid helium (LHe)
in pressurized storage dewars
to an altitude of 39.7~km.
Once at float altitude,
liquid nitrogen transfer
cooled a separate, unpressurized bucket dewar
to a temperature of 65~K,
followed by the transfer of 32 liters of liquid helium
from the storage dewar into the bucket dewar.
Calorimetric tests measured the total heat leak
to the LHe bath within bucket dewar.
A subsequent flight will replace 
the receiving bucket dewar
with an ultra-light dewar of similar size
to compare the performance of the ultra-light design
to conventional superinsulated dewars.
\end{abstract}

\maketitle


\begin{figure}[b]
\centerline{
\includegraphics[height=3.0in]{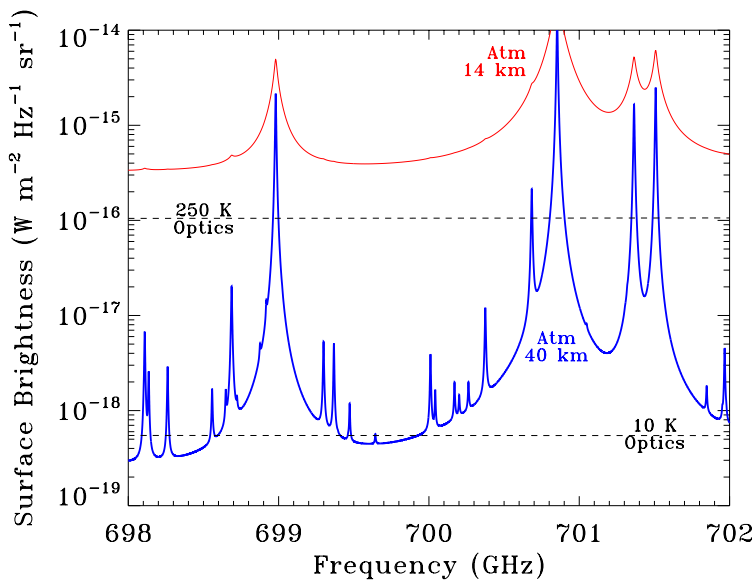} }
\caption{
Atmospheric emission dominates the diffuse sky brightness
even from airborne platforms such as SOFIA (thin red curve).
Observations from a balloon platform (thick blue curve)
reduce atmospheric emission by 2-3 orders of magnitude,
but require cryogenic optics 
in order to take advantage of the lower emission.}
\label{atm_fig}
\end{figure}

\section{Introduction}
Cryogenic telescopes flown at altitudes above 30~km
offer compelling astrophysical returns.
By reducing emission from both the atmosphere
and warm elements of the optics,
a cryogenic telescope
can improve sensitivity to astrophysical sources
by over two orders of magnitude
compared to observations from
ground-based or aircraft platforms.
Achieving such sensitivity gains
requires fully cryogenic optics.
Thermal emission from optical elements at 250~K
is 300 times brighter than the deepest atmospheric windows
(Figure \ref{atm_fig}).
Reducing emission from the telescope
to levels below the atmospheric emission
requires maintaining the beam-forming optics
at temperatures 10~K or colder.
This in turn presents a problem for balloon payloads.
Obtaining sufficient signal
and/or angular resolution
requires a large collecting area
({\it cf Herschel's} 3.5~m primary mirror diameter).
A dewar large enough to accommodate 
such large optics
would exceed balloon lift capability.
For example,
the ARCADE-2 and PIPER missions
flew cryogenic telescopes with 1~m clear aperture
\cite{singal/etal:2011,
pawlyk/etal:2018}.
The dewar alone weighed over 800 kg.
Scaled to 3~m aperture,
the resulting payload mass
of 5000 kg
easily exceeds 
the 2800 kg balloon lift capability.

The Balloon-Borne Cryogenic Telescope Testbed (BOBCAT)
develops technology
for large, lightweight dewars
at balloon altitudes.
We describe the BOBCAT program
and provide results
from the first of several planned demonstration flights.

\section{BOBCAT Mission}

A dewar has two functional requirements.
It must store a volume of cryogen
and
isolate the cryogen
from environmental heat.
The heat load is typically minimized
by surrounding the storage volume
with multiple layers of reflective material
(superinsulation)
maintained in a vacuum
to eliminate gas conduction.
Conventional superinsulated dewars
routinely achieve an areal heat load
below 1~W~m$^{-2}$.

\begin{figure}[b]
\centerline{
\includegraphics[width=3.0in]{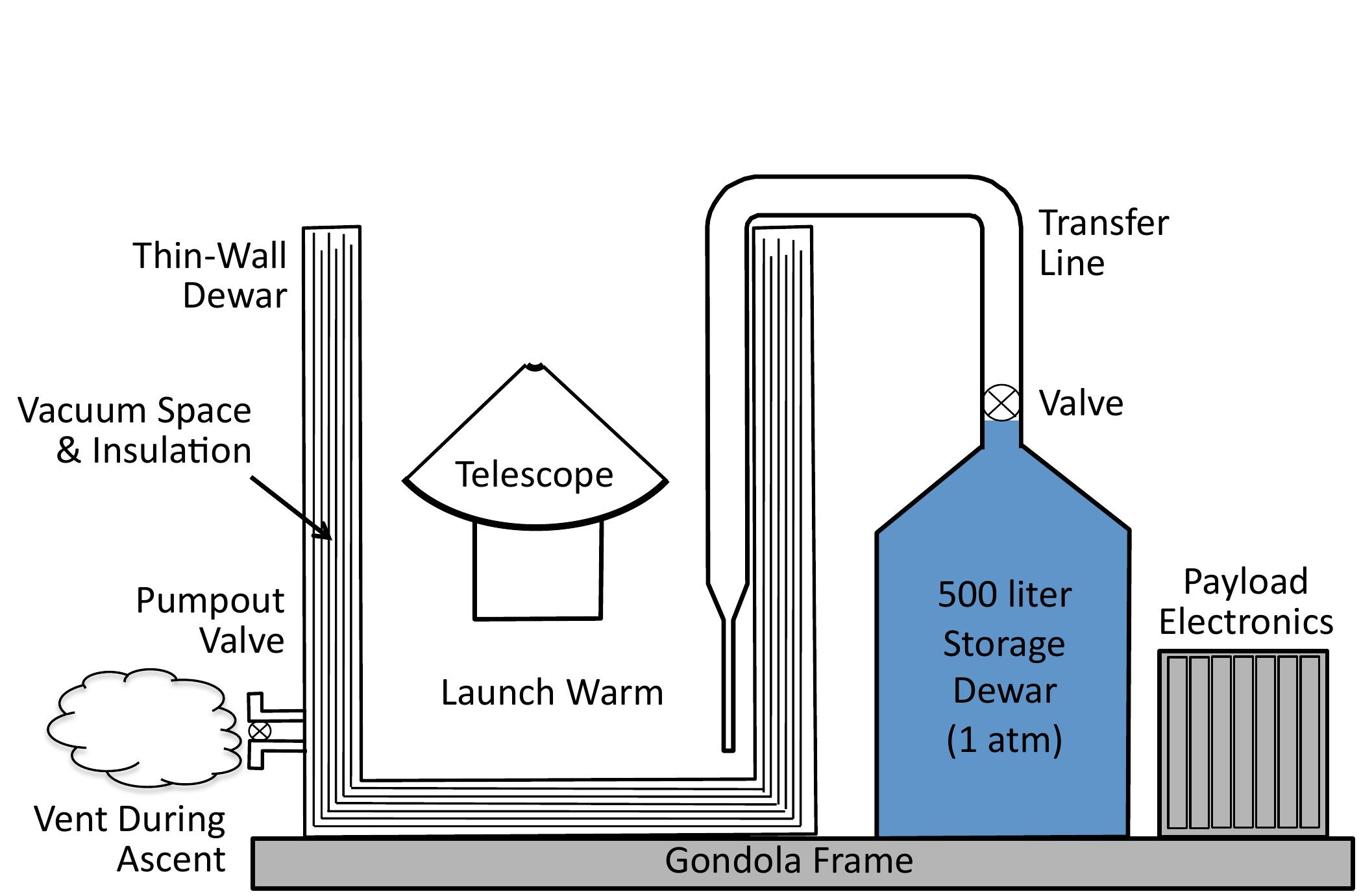} 
\includegraphics[width=3.0in]{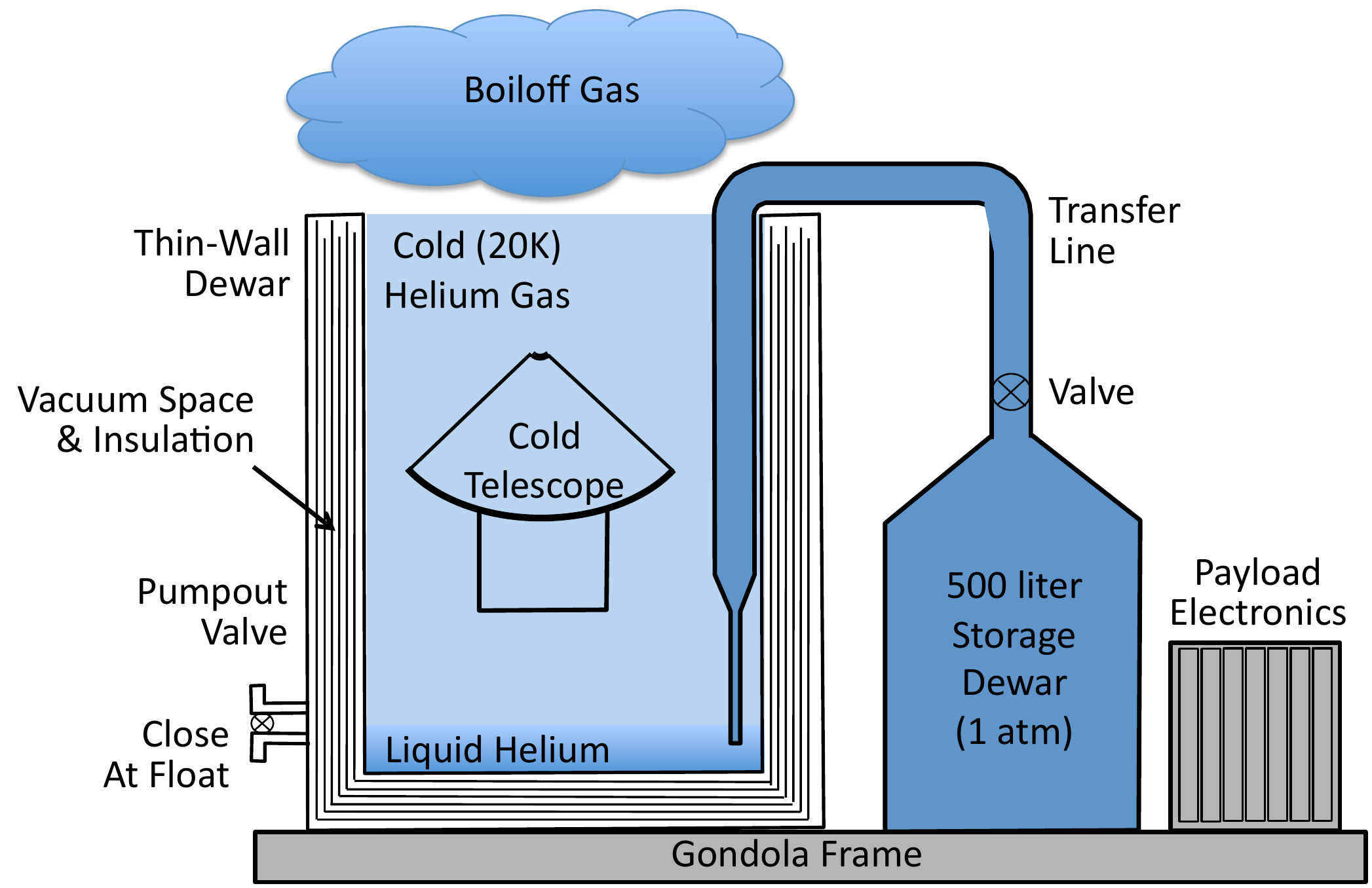} 
}
\caption{
Concept of the BOBCAT mission. 
A thin-wall bucket dewar launches empty (left panel) 
with the vacuum space vented 
so there is no pressure gradient across the walls.  
Once at float altitude (right) 
the vent is closed to isolate the vacuum space.
Cryogens are then transferred from a compact storage dewar
into the ultra-light dewar. 
The dewar walls need only be thick enough
to support the small pressure gradient at float altitude,
allowing significant mass savings.
Dewars as large as 4--5~m diameter could be flown
with existing balloon lift capability.
}
\label{bobcat_schematic}
\end{figure}

The dewar mass is dominated by the pressure walls
required to maintain the superinsulation
in vacuum.
The pressure walls must be thick to prevent catastrophic buckling
against the one-atmosphere pressure gradient at sea level.
The ambient pressure at 35 km altitude,
though,
is less than 0.3\% of the pressure at sea level.
A thin-wall dewar optimized for operation at balloon altitudes
could achieve significant mass savings
provided that it never hold cryogen at sea level.
Figure \ref{bobcat_schematic} shows the concept.
BOBCAT consists of a dewar with thin (0.5~mm) stainless steel walls. 
The dewar launches warm 
with its vacuum space vented continuously during ascent, 
thereby eliminating any pressure gradient across the walls. 
Once float altitude is reached, 
the valve is closed to seal the ultra-light dewar's vacuum space. 
A separate (standard construction) storage dewar 
maintains a reservoir of liquid helium
with minimal parasitic mass.  
It launches with the payload, 
connected to the ultra-light dewar by a vacuum-insulated transfer line.
Once the valve closes to seal off the vacuum space,
cryogens (LN2 and LHe)
may be transferred from the storage dewar
into the ultra-light dewar.
Cryopumping of the residual gas within the vacuum space
efficiently eliminates direct gas conduction
between the radiative shields,
allowing the ultra-light dewar to function normally.
The cold boiloff gas from the ultra-light bucket dewar
forms a barrier between the cryogenic optics and the atmosphere: 
there are no warm windows to overwhelm the sky signal. 
As demonstrated by ARCADE, 
helium gas below 20~K is denser than the ambient atmosphere at float
and forms a stable ``pool'' within the bucket dewar
to prevent nitrogen condensation on the cold optics below\cite{
singal/etal:2011,
fixsen/etal:2011}.

A key requirement is the ability to efficiently transfer
large volumes of cryogenic liquid
from a storage dewar 
to the ultra-light dewar
while at float altitude.
Pressure gradients
between the storage dewar
and the ambient environment
provide a simple means for such transfer,
requiring only a remotely-controllable valve
to initiate and terminate cryogen flow.
Unlike the Superfluid Helium On-Orbit Transfer mission\cite{
dipirro/castles:1986},
which demonstrated superfluid LHe transfer 
between two vessels both maintained at near-vacuum pressure,
the gradient between the
high-pressure storage dewar 
and low-pressure ultra-light receiving dewar
presents two potential complications.
Liquid nitrogen is commonly used
to pre-cool instrumentation
prior to helium transfer.
The 1--2~Torr ambient pressure at float altitude
is well below the 94~Torr triple point of nitrogen.
Liquid nitrogen exiting the transfer line
will thus freeze into a nitrogen ``snow,''
potentially clogging the transfer line.
Liquid helium undergoes a superfluid phase transition
as the pressure falls below 38~Torr.
Back pressure from the resulting boiloff 
can create standing waves
(thermal-acoustic oscillations)
within the transfer line
to reduce or stop liquid transfer.

The BOBCAT program develops technology 
for a future balloon-borne observatory.
It validates the required fabrication and operational techniques
while 
demonstrating key performance parameters
(heat leak, hold time)
for a thin-walled bucket dewar.
It proceeds in several phases.
A first phase (BOBCAT-1)
flew in August 2019
to demonstrate the necessary cryogen transfer and flow control at altitude
using a standard superinsulated dewar with 25~cm aperture.
A second phase (BOBCAT-2)
will fly at a later date,
replacing the standard superinsulated dewar
with an ultra-light dewar
of identical size
to demonstrate the thermal performance of the ultra-light concept.
For simplicity,
both BOBCAT-1 and BOBCAT-2
seal the top of the receiving bucket dewar with a foam lid
to isolate the liquid space
from the ambient atmosphere;
there is no ``pool'' of cold helium gas
for either flight.
Additional flights
will characterize
the helium gas pool above an open-aperture dewar
to quantify the astronomical seeing through the gas column.
Results from the full program can then inform the design
and operation of future large (3--5~m aperture) cryogenic missions
at balloon altitudes.

\begin{figure}[b]
\centerline{
\includegraphics[width=4.0in]{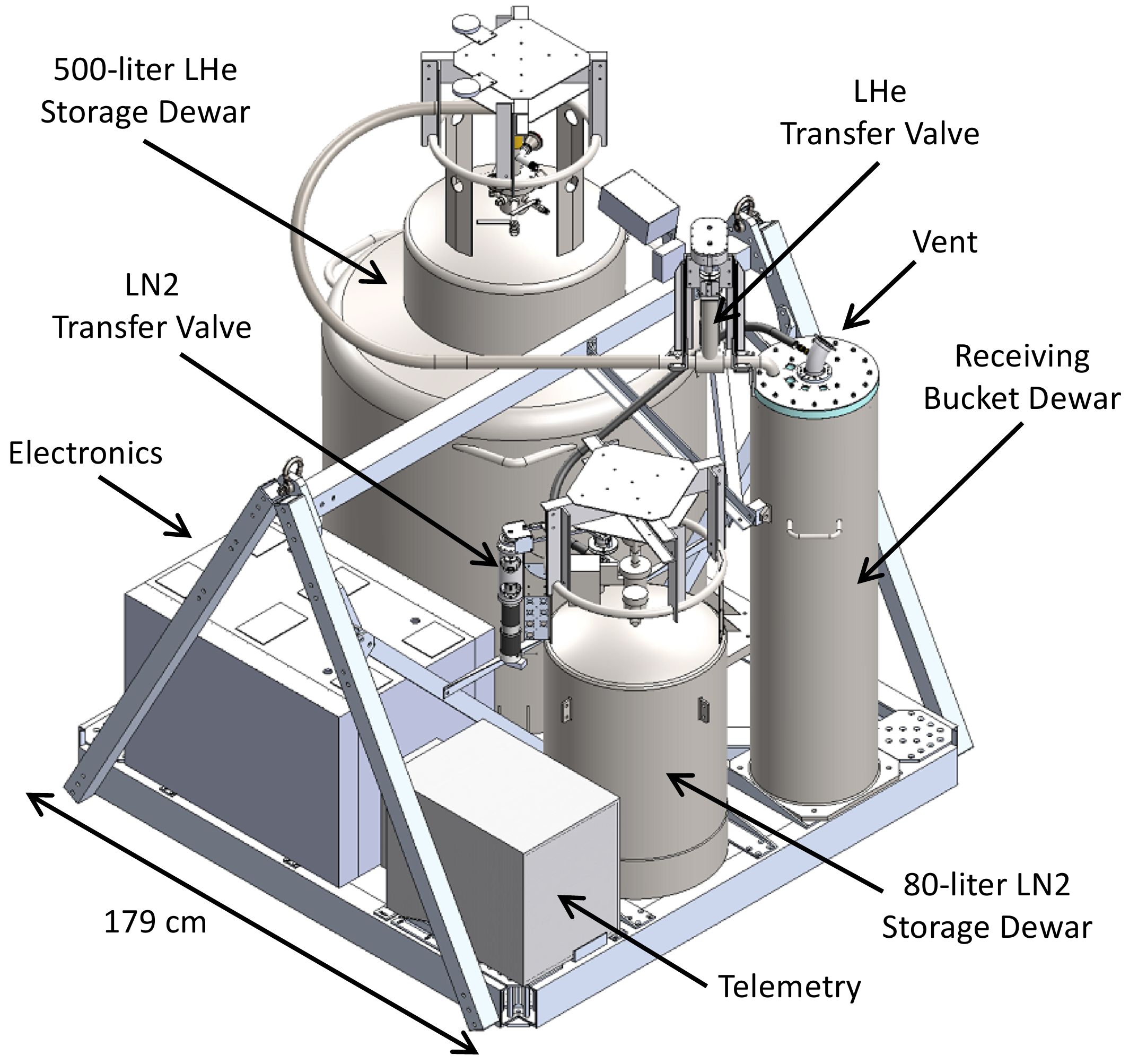} 
}
\caption{
Schematic drawing of the BOBCAT-1 payload.
}
\label{bobcat_gondola}
\end{figure}

\section{BOBCAT-1 Instrumentation}

Figure \ref{bobcat_gondola} shows the BOBCAT-1 payload.
It consists of a 500-liter LHe storage dewar, 
an 80-liter LN2 dewar, 
and an electronics module on a simple aluminum gondola frame.  
A conventional superinsulated stainless-steel bucket dewar
with liquid storage volume 59 liters
stands in for the ultra-light dewar
and receives cryogens from the two storage dewars.
A video camera provides real-time visual feedback. 
Telemetry boards copied from PIPER
\cite{
pawlyk/etal:2018,
lazear/etal:2014}
provide temperature and pressure readout 
from multiple locations.
Table 1 summarizes the mass budget for the BOBCAT-1 payload.
%
%

The payload includes 2 conventional storage dewars,
modified for use on a high-altitude payload.
An 80-liter dewar
(Cryofab model CLPB-80)
holds liquid nitrogen to pre-cool the test dewar.
It has been modified
to include 
an American Magnetics capacitive level sensor,
a 50~W pressure builder,
and a separate telemetry port
for pressure sensors.
A 500-liter dewar
(International Cryogenics model IC-500)
holds liquid helium.
It has been modified
to replace the standard neck assembly with a
conflat flange,
which contains
a liquid withdrawal line,
vent line,
temperature and pressure sensors,
two redundant American Magnetics superconducting level sensors,
and two redundant 52~W heaters / pressure builders
which also allow rapid boiloff of residual cryogen
at the end of the flight.
Custom-installed pressure relief valves on each storage dewar
maintain the cryogen at
one atmosphere above ambient pressure,
allowing each dewar to vent as needed during ascent.

\begin{table}[t]
{
\small
\caption{Payload Mass Budget}
\label{fringe_table}
\begin{center}
\begin{tabular}{l r}
\hline 
Subsystem		&	Dry Mass \\
			&	(kg)	 \\
\hline
LN2 Storage Dewar	&	75	 \\
LHe Storage Dewar	&	320	 \\
LN2 Transfer Line	&	3	 \\
LHe Transfer Line	&	20	 \\
Bucket Dewar		&	70	 \\
Avionics		&	114	 \\
Telemetry		&	30	 \\
Frame			&	150	 \\
\hline
Total			&	782	 \\
\hline
\end{tabular}
\end{center}
}
\end{table}

A standard stainless steel transfer line
(McMaster-Carr model 54935K61)
connects the LN2 storage dewar to the receiving test dewar.
A DC gearmotor 
(Midwest Motion Products model MMP D22-376H-24V GP52-1140 E5-032)   	
couples to the handle of the storage dewar's liquid withdrawal valve
to control LN2 flow to the test dewar.
A vacuum-insulated transfer line
(International Cryogenics model 92-9552)
with 4.77~mm inner diameter
connects the LHe storage dewar to the receiving dewar.
A separate, identical DC gearmotor
turns a cryogenic globe valve built into the helium transfer line
to control LHe flow to the test dewar.
A shaft encoder,
tachometer,
current monitor, and voltage monitor
provide real-time telemetry
for each valve motor.
Each motor includes a thermometer
and heater
for thermal control;
the LN2 valve also has a heater/thermometer pair.

\begin{figure}[b]
\centerline{
\includegraphics[height=3.0in]{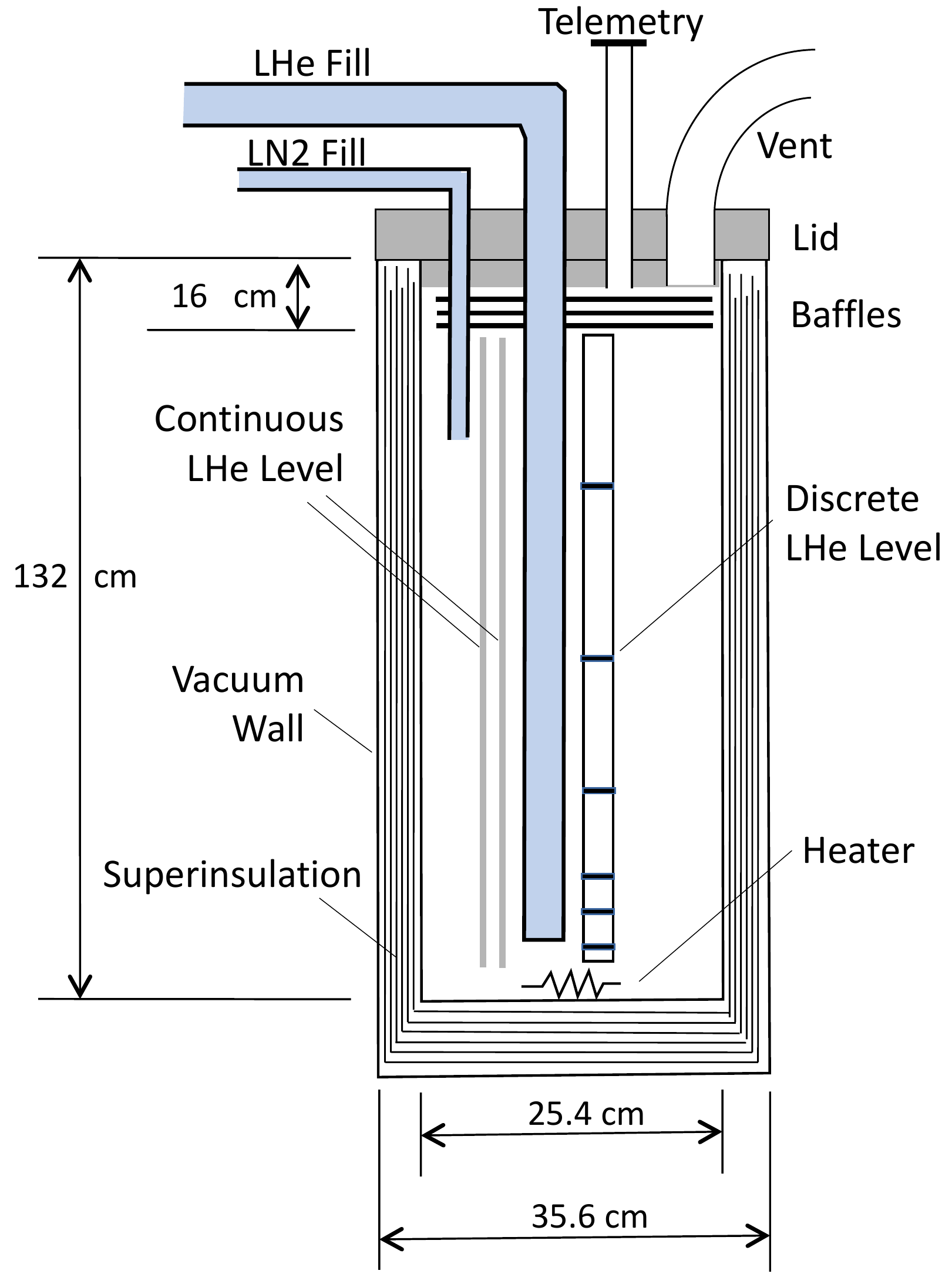} 
}
\caption{
Schematic drawing of the receiving dewar.
Temperatures
and liquid helium level
are monitored through a telemetry port.
}
\label{babybear_fig}
\end{figure}

Figure \ref{babybear_fig} shows the receiving dewar.
It consists of a conventional bucket dewar 132 cm tall
and 35.6 cm outer diameter,
using multi-layer superinsulation
within stainless steel vacuum walls
to enclose a 59-liter cylindrical liquid space with 25.4 cm diameter.
The vacuum space remains sealed throughout the flight.
A lid made from extruded polystyrene foam
(Dow blue Styrofoam\textsuperscript{TM} sheet)
prevents atmospheric ingress
and supports
the two cryogen fill lines,
a telemetry port,
and an open vent.
A set of 8 silicon diodes
(Lakeshore DT-670)
and 6 ruthenium oxide thermometers
(Lakeshore RX-103)
monitor temperatures
throughout the dewar liquid space.
Two redundant 
American Magnetics 
superconducting level sensors
91 cm tall
monitor the LHe level.
A discrete level sensor
mounts 
a set of heaters and resistive ruthenium oxide thermometers
at heights 
2.5, 7.5, 13, 23, 38, and 58 cm
above the dewar bottom.
When powered,
each 2~W heater
will heat the thermometer,
decreasing its resistance
from 38~k$\Omega$
to 11~k$\Omega$
if the thermometer is not immersed in LHe.
As with the LHe storage dewar,
separate heaters (14~W and 72~W)
on the dewar bottom
can be powered
to rapidly boil off any cryogens
remaining at the end of the flight,
with the two sizes intended
to provide fine and coarse control if needed.
\section{BOBCAT-1 Flight}

\begin{figure}[b]
\centerline{
\includegraphics[height=5.0in]{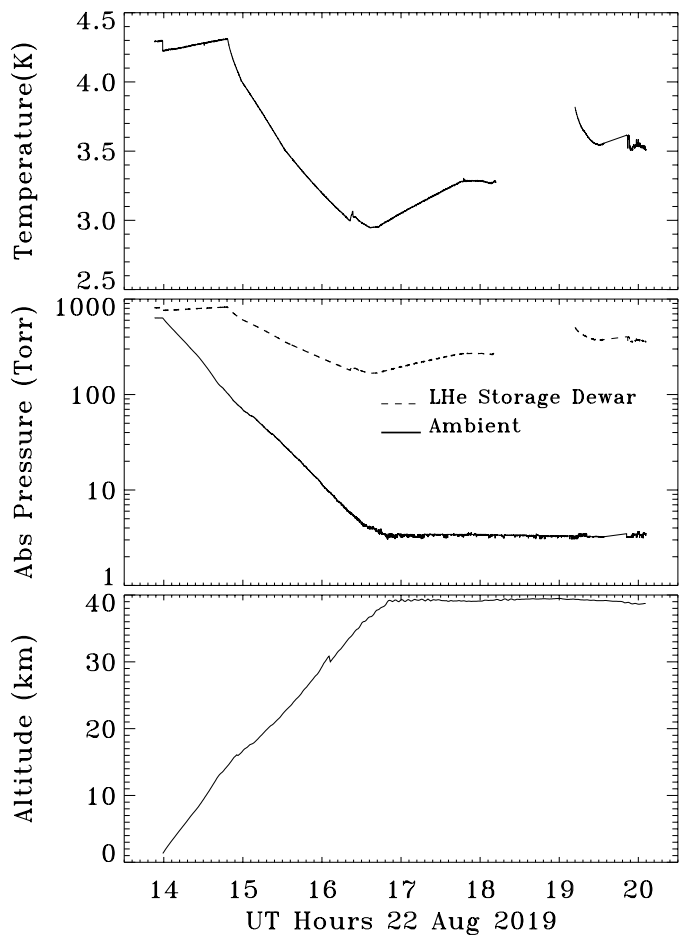} 
}
\caption{
Temperature and pressure within the 500-liter storage dewar
throughout flight.
The ambient pressure and payload altitude are also shown.
Telemetry within the storage dewar is unreliable during
active LHe transfer.
Pressure within the storage dewar fell below the 1 atmosphere design goal
but remained well above the superfluid transition.
}
\label{ascent_fig}
\end{figure}

BOBCAT-1 launched at 13:59 UT
on 22 Aug 2019
from NASA's Columbia Scientific Balloon Facility 
in Ft Sumner, NM
(elevation 1200 m)
carrying 14 liters of liquid nitrogen
and 268 liters of liquid helium.
The LHe storage dewar was vented to atmosphere 
several hours prior to launch
so that its internal pressure at launch (810 Torr)
was only modestly above the ambient 656 Torr.
A 0.84 million cubic meter balloon
carried the payload,
reaching float altitude of 39.7~km
at 16:52 UT.
During ascent,
temperatures within the 500 liter storage dewar 
remained steady until the ambient pressure
dropped below 80 Torr,
after which the temperature fell steadily
from 4.3~K to 3~K,
below the 4.2~K value
expected for the one-atmosphere overpressure relief valve
but
consistent with a slow leak in the neck assembly
or a stuck / frozen pressure relief valve.
Figure \ref{ascent_fig}
shows the pressure,
temperature,
and altitude during the flight.
At 16:20 UT
a 52~W heater within the 500-liter storage dewar was turned on
for 3 minutes
to increase the pressure
and maintain LHe temperature well above the
superfluid transition.
The temperature of the LHe bath within the dewar 
stabilized at 3.2~K
after the heater turned on,
consistent with a pressure of 300 Torr inside the dewar
compared to 1.2 Torr ambient pressure at float.

\begin{figure}[b]
\centerline{
\includegraphics[height=5.0in]{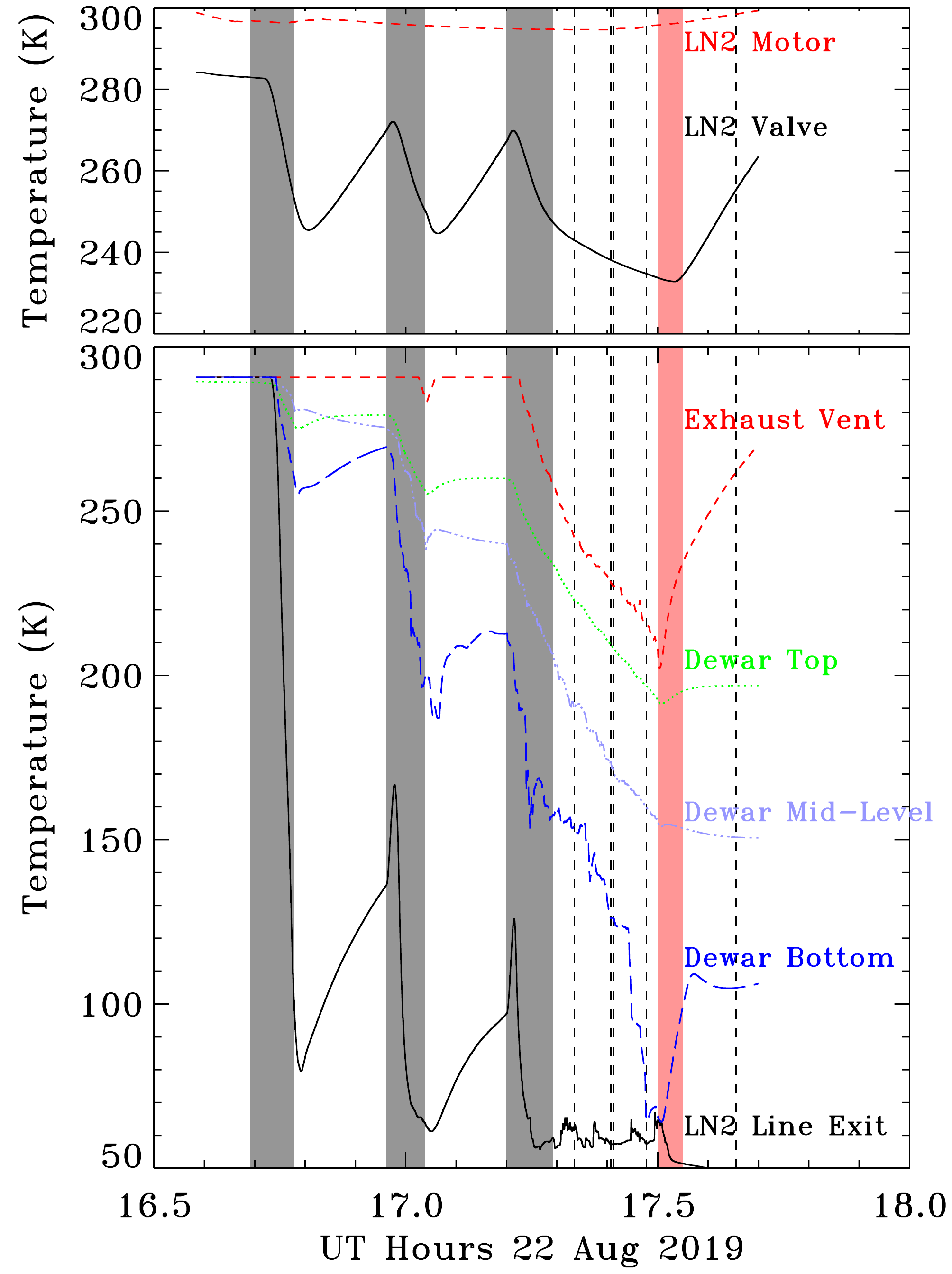} 
}
\caption{
Selected temperatures during the LN2 pre-cooling operation.
The top panel shows
the LN2 valve and motor temperatures
while the bottom panel shows 
shows temperatures within the receiving bucket dewar.
Vertical gray bars show periods when the LN2 valve was commanded open.
Falling temperatures after the third and final LN2 transfer
indicate a stuck or frozen valve.
Dashed vertical lines indicate additional commands to close the valve
while the vertical red bar shows the period 
with the two boiloff heaters turned on.
}
\label{ln2_precool_fig}
\end{figure}

Cryogen transfer began at 16:38 UT
using LN2 to pre-cool the receiving dewar.
To minimize collection of frozen nitrogen
within the receiving dewar,
the LN2 liquid withdrawal valve 
was only opened to 6\% of its full-open position
({\it i.e.} barely open),
using the 1000~Torr pressure gradient between the LN2 storage dewar
and the receiving dewar 
to force LN2 through the transfer line.
The valve opened three times for a six-minute duration each time,
and then closed for ten minutes 
to allow temperatures within the receiving dewar
to stabilize before opening the valve again.
Figure \ref{ln2_precool_fig} shows selected temperatures
during the nitrogen precooling operation.
During each nitrogen transfer,
the LN2 liquid withdrawal valve 
cooled rapidly,
reaching temperatures below 260~K
before the valve was closed each time.
A 24~W heater on the valve 
raised the temperature
once the valve closed,
but proved unable to maintain a constant temperature
during nitrogen transfer.
During the third and final nitrogen transfer,
the valve temperature continued to fall
after the valve was commanded to the closed position.
Temperatures within the receiving dewar
also continued to fall,
consistent with
the continued flow of nitrogen through the valve
into the receiving dewar.
It is probable that frozen nitrogen 
held the valve slightly open
despite several commands to close.
By 17:30 UT the bottom of the receiving dewar
reached 63~K,
indicating a layer of frozen nitrogen.
Both boiloff heaters on the dewar bottom
were then turned on for 3 minutes
to sublimate the frozen nitrogen,
after which temperatures at the dewar bottom stabilized near 100~K.
Once the boiloff heaters turned on,
the LN2 valve temperature recovered
and we were able to fully close the valve.

\begin{figure}[b]
\centerline{
\includegraphics[height=3.0in]{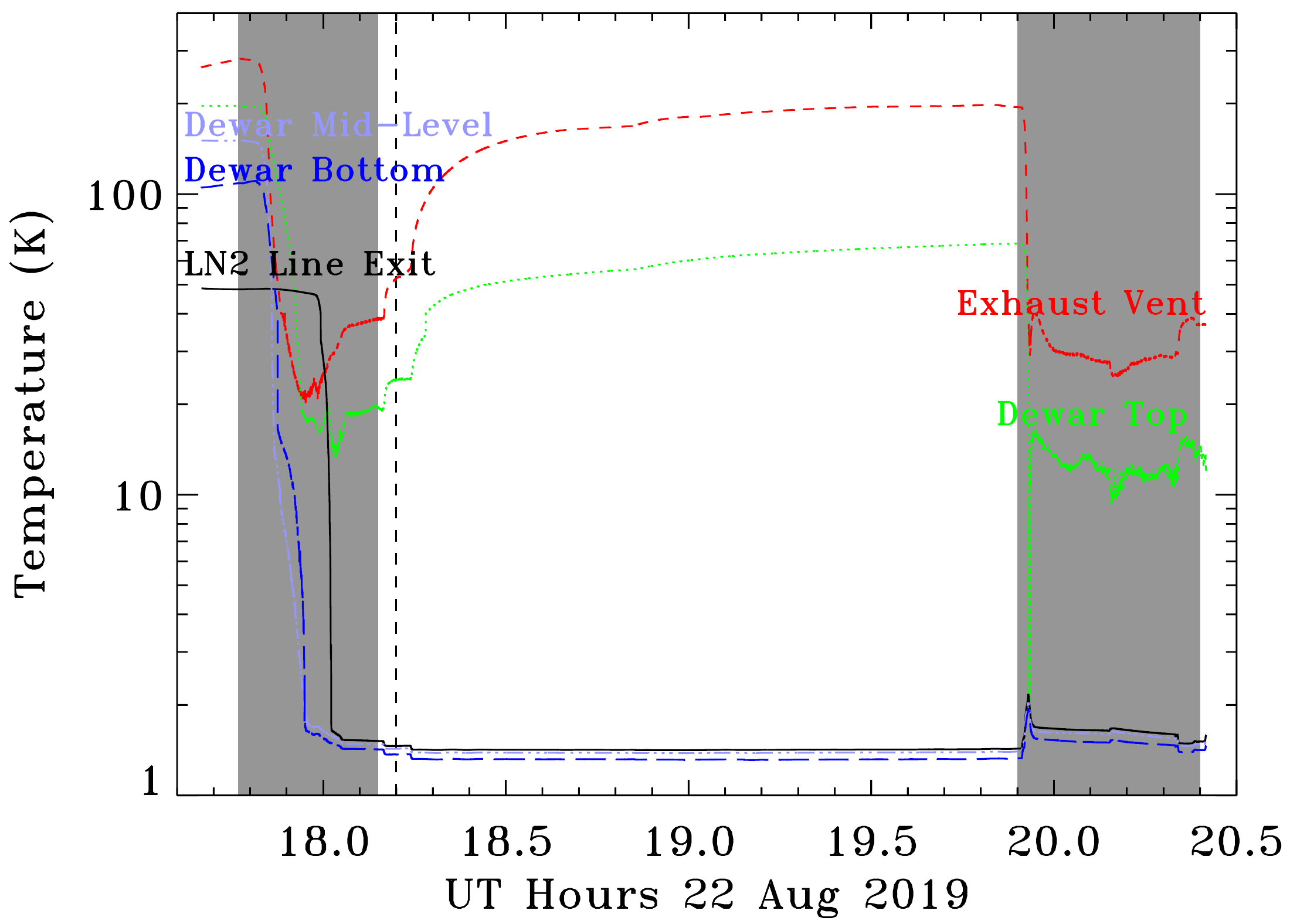} 
}
\caption{
Selected temperatures within the receiving bucket dewar
during during LHe operations.
The vertical gray bars show the periods
when the LHe valve was commanded open.
The dashed vertical line indicates an additional command 
to close the valve.
}
\label{lhe_cooldown_temps}
\end{figure}

Liquid helium transfer began at 17:46 UT.
To achieve some additional cooling from the helium gas enthalpy,
the LHe valve was only commanded to 15\% of the full open position,
minimizing the initial transfer rate.
Figure \ref{lhe_cooldown_temps} shows temperatures within the
receiving bucket dewar during LHe operations.
Temperatures within the receiving dewar
fell rapidly,
reaching the LHe boiling temperature of 1.3~K
at 17:57 UT.
By 18:09 UT
the receiving dewar held 32 liters of LHe,
at which point the valve on the LHe transfer line was closed
and cryogen transfer terminated.

\begin{figure}[b]
\centerline{
\includegraphics[height=4.0in]{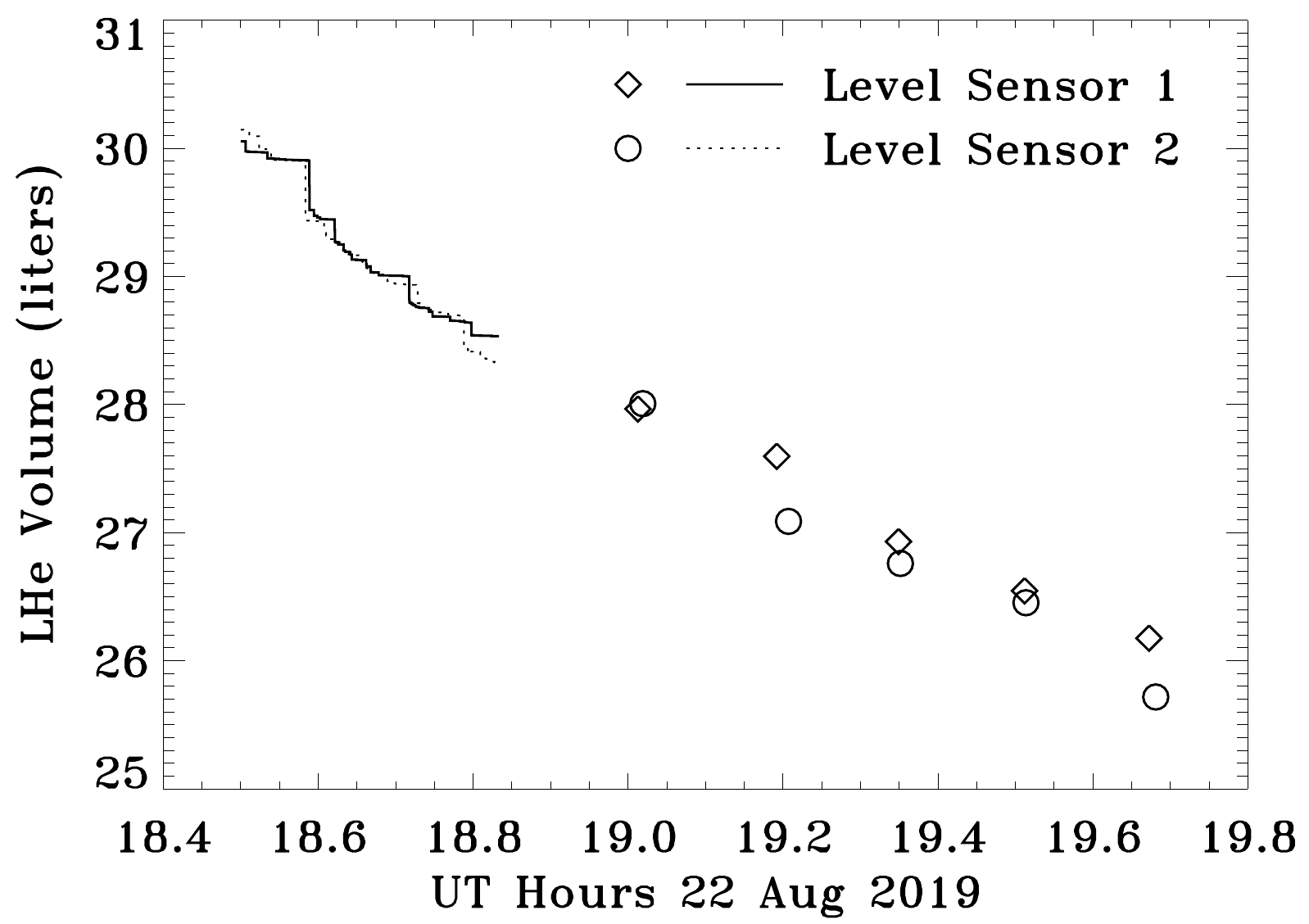} }
\caption{
Liquid helium volume within the receiving dewar
during passive calorimetric testing.
Two redundant level sensors 
monitored the liquid level continuously 
for the first half of the test,
then were intermittently powered during the second half
to minimize their dissipated power.
}
\label{calorimetric_fig}
\end{figure}

The primary goal of the 2019 flight
was a demonstration of cryogen transfer
at float altitude.
A secondary goal was the calorimetric measurement
of the total heat leak to the receiving dewar,
which will be compared in a follow-up flight
to a similar measurement using an ultra-light receiving dewar
of identical dimensions.
This test proceeded in two stages.
We paused for 20 minutes after the LHe transfer
to allow temperatures to settle in the receiving dewar,
after which
both continuous level sensors were turned on.
Figure \ref{calorimetric_fig}
shows the subsequent LHe volume as a function of time.
From 18:30 to 18:50 UT
we measure a LHe loss rate of $5.4 \pm 0.2$~l~hr$^{-1}$,
corresponding to a total heat leak of $4.7 \pm 0.2$~W.
The power dissipated by the level sensors
depends on the liquid level.
We monitor the  current and voltage in each sensor
to correct for the dissipated power.
During the continuous level test,
sensor 1 dissipated average power 0.76~W
while sensor 2 dissipated average power 1.07~W.
Correcting for the LHe loss due to sensor power dissipation
yields a corrected heat leak of $2.8 \pm 0.2$~W
for the dewar. 

The second half of the test
monitors the LHe level intermittently,
limiting the duty cycle of the two level sensors
to reduce their dissipated power.
From 19:00 to 19:40 UT
each sensor was powered 5 times
for approximately 40 seconds each time,
for a total combined dissipation of 285 joules.
After correction for this minor dissipation,
the measured loss rate of
$3.2 \pm 0.3$~l~hr$^{-1}$
corresponds to a heat leak
$2.7 \pm 0.2$~W.

\begin{figure}[b]
\centerline{
\includegraphics[height=4.0in]{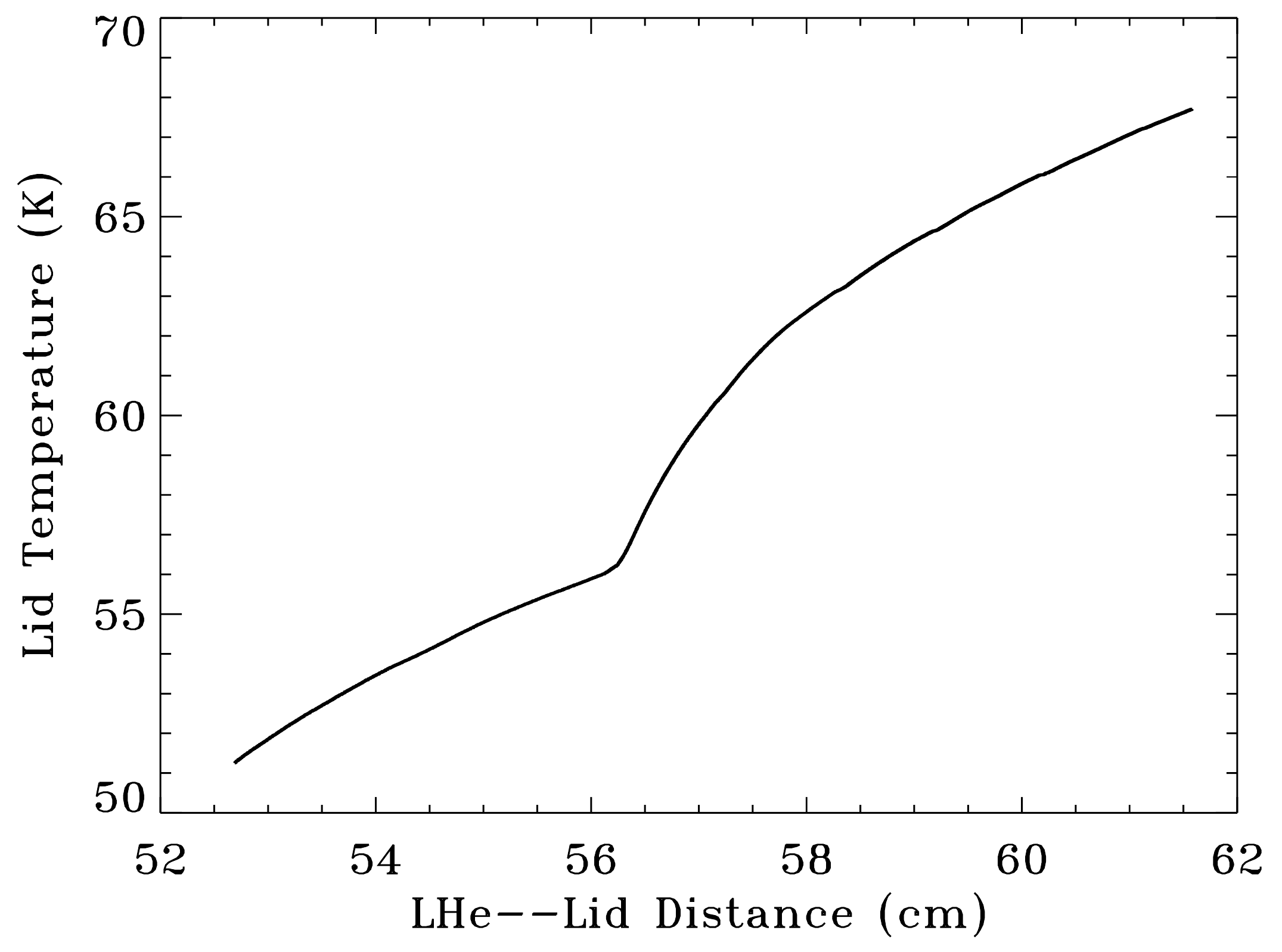} 
}
\caption{
The temperature underneath the receiving dewar lid
during calorimetric testing
is shown as a function of the distance to the liquid below.
The reduced duty cycle of the continuous LHe level sensors
during the second half of the test
causes the break in slope at distances above 56~cm.
Heat flow from the lid to the liquid
depends on the temperature gradient and the lid--liquid distance;
the higher temperatures are compensated by longer distances
so that the conducted heat during calorimetric testing
is constant within 11\%.
}
\label{stratification_fig}
\end{figure}

The calorimetric tests are both consistent
with a heat leak of  $2.7 \pm 0.2$~W
to the interior of the receiving dewar.
This includes heat conduction through 
the 1.2~m$^2$ superinsulation
as well as
conduction down the interior wall
and along the insert
holding the transfer lines and level sensors.
Stratification of boiloff gas within the dewar
during these tests
could affect the latter two terms,
but the effect should be nearly constant
over the limited change of LHe level during calorimetric testing.
Figure \ref{stratification_fig}
shows the temperature at the top of the dewar
(just beneath the lid)
during calorimetric testing,
as a function of the distance to the liquid below.
Conductive heat flow from the top of the dewar to the liquid
should scale linearly with the temperature gradient
and inversely with the lid--liquid distance.
Throughout calorimetric testing,
the ratio $\Delta T/L$ varies by only 11\%.

At 19:54 UT preparations began to terminate the flight.
The LHe valve was opened again to remove helium
from the 500-liter storage dewar,
transferring an additional 20 liters
to the receiving dewar.
At the same time, heaters 
in both the 500-liter storage dewar
and the receiving dewar
were turned on to boil off all remaining cryogen.
By 20:28 UT both LHe dewars were empty.
The nitrogen storage dewar was not emptied,
and was left with 11 liters LN2
at one atmosphere overpressure.

Pressure within the 500-liter storage dewar
remained at 380 Torr,
well below both the one-atmosphere design
and the ambient pressure at landing.
To prevent potential buckling of the storage dewar 
from internal underpressure during descent,
the LHe valve between the 500-liter storage dewar
and the receiving dewar was left in the full open position
so that the storage dewar
would remain at ambient pressure throughout descent.
The flight ended at 21:28 UT;
the payload was subsequently recovered undamaged.

\begin{figure}[b]
\centerline{
\includegraphics[width=5.0in]{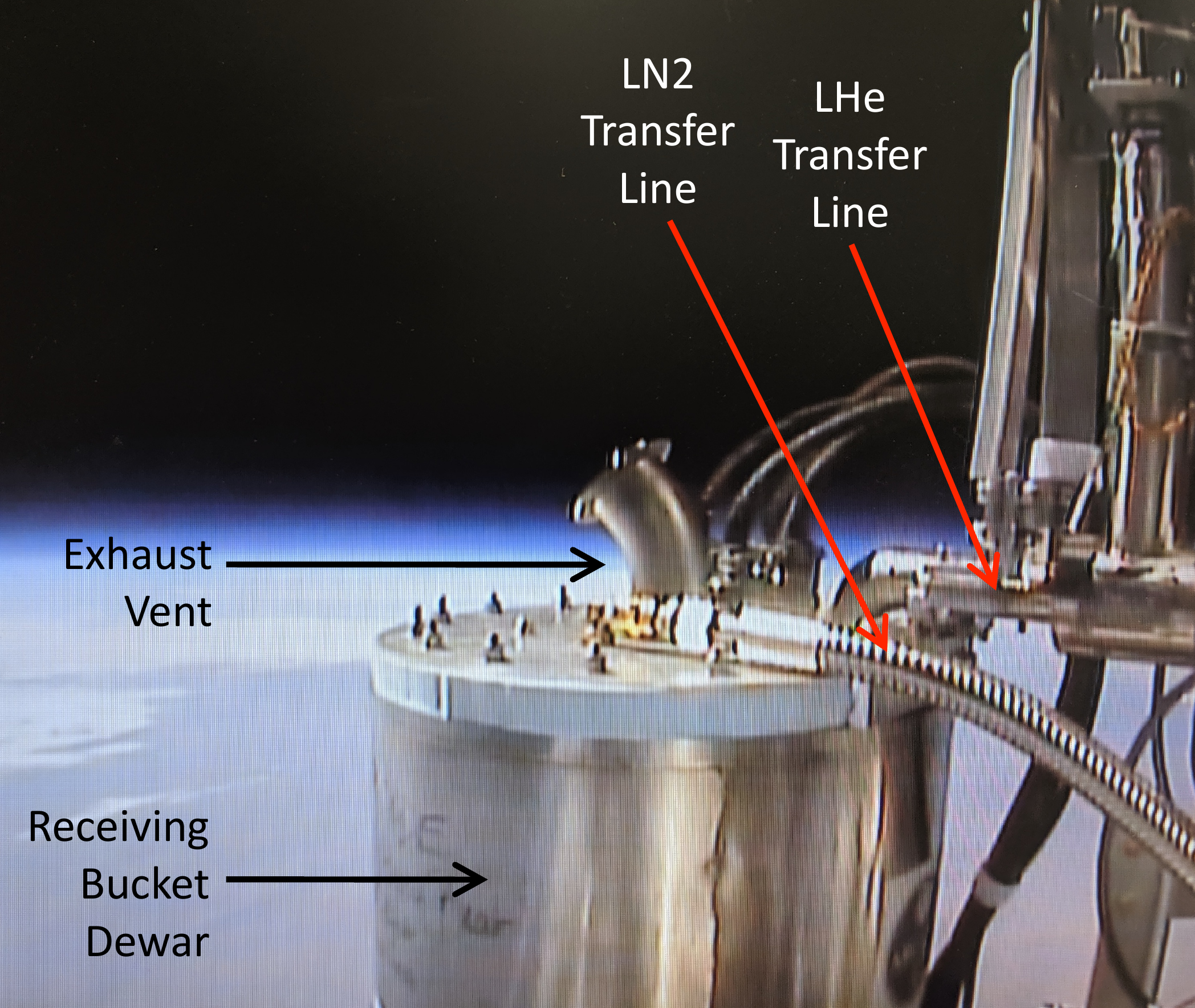} 
}
\caption{
In-flight video still showing the receiving dewar exhaust vent
during LHe transfer.
Although the vent is at 30~K,
no condensation plume is visible.
}
\label{video_still}
\end{figure}

The BOBCAT instrument package included a real-time video link
with the camera focused on the vent line 
exiting the receiving dewar.
At no time throughout the flight 
was any plume of condensation visible.
This includes the initial LN2 cooldown,
the LHe transfer for calorimetric testing,
and the final LHe transfer and boiloff
prior to termination.
Figure \ref{video_still} shows the dewar and exhaust vent
during the initial LHe fill.
Both the initial LHe fill
and the final boiloff 
reduced the temperature of the vent exhaust gas below 30~K
but produced no visible plume into the ambient atmosphere.
Additional instrumentation
to determine the extent (if any)
of atmospheric condensation
from cryogenic operations at 39~km altitude
is planned for a future flight.

\section{Discussion}

Cryogen expenditure and transfer efficiencies 
during BOBCAT-1 flight agreed with
pre-flight ground tests at Ft Sumner.
The nitrogen transfer at float consumed 3 liters of LN2
while cooling the bucket dewar liquid volume
from initial temperature 297~K
to 100~K at the dewar bottom
and 200~K at the top.
Ground tests also required 3 liters of LN2 to achieve similar cooling. 
The 500-liter storage dewar
launched with 268 liters LHe,
of which 168 liters remained once float altitude was reached.
Cooling the bucket dewar to 1.3~K
and transferring 32 liters LHe
required a total LHe usage of 61 liters,
including the losses to the superfluid phase transition.
By comparison,
ground tests required 35 liters LHe
to transfer 9 liters into the bucket dewar.
Once the transfer line and receiving dewar 
are cooled to LHe temperatures,
additional parasitic losses 
are a small fraction of the transferred helium volume.
Ground tests thus required some 26 liters of LHe
to cool the transfer line and receiving dewar,
compared to 29 liters required at float.
The measured values agree with the estimated LHe volume
needed to cool the $\sim$10~kg of stainless steel
in the transfer line and dewar inner walls.
The helium volume required to cool future missions
using much larger dewars
can thus be estimated 
from the total thermal mass to be cooled.

The total heat leak of $2.7 \pm 0.2$~W measured in flight
is larger than the 1--2~W measured 
using the same dewar in ideal laboratory configurations.
From the measured top-bottom temperature gradient
(Fig \ref{stratification_fig})
the parasitic heat flow down the dewar wall
will be of order 0.4~W
(neglecting cooling through the gas)
and should be comparable for both ground and flight operations.
Laboratory configurations include a longer (20~cm) 
set of radiative baffles between the lid and the liquid,
and did not include the fixed transfer lines
and sensor suite.
Sensors within the storage dewars and receiving bucket dewar 
are designed for ruggedness,
not to minimize parasitic heat leaks.  
In particular, the copper wire leads 
to the thermometers and heaters inside the dewars 
have diameters of 0.32~mm and 0.80~mm, respectively.
Prior experience with balloon payloads has shown that
the 0.13~mm wire more typical of laboratory dewars
tended to develop shorts or breaks
during cross-country shipping to the launch site.
Thermal conduction through the wires
could contribute as much as 0.65~W during calorimetric tests
if cooling through the gas column can be neglected.
Solar heating of the dewar during flight
creates an additional potential source of heat.
The BOBCAT-1 flight took place in daylight.
The temperature of the unpainted stainless steel dewar walls
was not recorded during flight,
but nearby unpainted aluminum boxes
warmed to 342~K, well above the 300~K temperatures
typical for ground testing.
If the dewar exterior wall reached comparable temperatures,
heat flow through the dewar superinsulation
would be expected to increase by 10--50\%,
depending on the ratio of conductive to radiative transport
within the superinsulation.
Other sources of heat in flight are small.
Direct thermal radiative from the 70~K lid
contributes less than 60 mW to the total load.

The BOBCAT project develops technology and operational techniques
for a future large (3--4 meter) cryogenic telescope
on a balloon platform.
The BOBCAT-1 flight provides a baseline to compare the performance
of a standard-construction superinsulated dewar to an ultra-light
dewar of comparable dimensions.  
A second flight will compare the cryogenic performance
of an ultra-light dewar to 
the standard-construction superinsulated dewar described here.
The liquid space of the BOBCAT-2 ultra-light dewar 
will have identical dimensions
as the BOBCAT-1 superinsulated dewar
and will re-fly the same transfer lines,
lid assembly, and sensor suite
so that any differences in the calorimetric heat leak
between the BOBCAT-1 and BOBCAT-2 flights
can be attributed to the dewar design
and not the instrumentation.

The BOBCAT-1 flight shows that bulk cryogen transfer
at float altitude
can proceed 
on time scales 
and with cryogen losses 
comparable to ground operations,
despite the low operating pressure at float.
We find no significant differences
between flight and ground tests
for the cryogen volume 
required to cool the dewar and instrument package,
or for the time required for cryogen transfer.
Several simple modifications to the BOBCAT-1 payload
would facilitate future operations.
The heater on the nitrogen valve was unable 
to maintain stable temperatures during active LN2 transfer.
While a larger heater could mitigate this problem,
use of a custom LN2 transfer line 
with a built-in valve 
(comparable to the LHe transfer line and valve)
would provide physical separation between the cryogen flow
and the valve motor,
eliminating the problem.
The largest cryogen loss 
from the low operating pressure 
(a 30\% loss of helium mass from the storage dewar
at the superfluid transition)
could be avoided in future large missions
using a storage dewar pumped below the superfluid transition
prior to flight.
Use of a pumped storage dewar would additionally eliminate
the need to backfill the dewar prior to descent,
removing the possibility of ice plugs forming
in the storage dewar during descent.

\begin{acknowledgments}
We thank the CSBF staff for launch support
and 
gratefully acknowledge P. Cursey,
N. Gandilo,
S. Pawlyk,
and P. Taraschi
for their contributions.
This work has been funded by the IRAD program
at NASA's Goddard Space Flight Center.
\end{acknowledgments}

\vspace{3mm}
{\bf
{\small
DATA AVAILABILITY STATEMENT}} 

The data that support the findings of this study 
are available from the corresponding author upon reasonable request.

\bibliography{bobcat1_xfer_rsi}		

\end{document}